\documentclass[12pt,preprint]{aastex}
\usepackage{natbib}
\usepackage{graphicx}
\usepackage{amsmath}
\usepackage{epsfig}
\usepackage{graphicx}
\usepackage{rotating}
\usepackage{lscape}
\newcommand{\asec}{$^{\prime\prime}$}

\def\H{N$_{2}$H$^{+}$}
\def\D{N$_{2}$D$^{+}$}

\def\AMM{NH$_3$}

\def\METH{CH$_3$OH}

\def\HII{H{\sc ii}}

\def\kms{\mbox{km~s$^{-1}$}}
\def\cmc{cm$^{-3}$}
\def\cmq{cm$^{-2}$}

\def\Tex{\mbox{$T_{\rm ex}$}}

\begin{document}
\title{Phosphorus-bearing molecules in massive dense cores
\thanks{Based on observations carried out with the IRAM-30m Telescope. IRAM is supported by INSU/CNRS (France), MPG (Germany) and IGN (Spain).}}
\author{F. Fontani$^{1}$, V.M. Rivilla$^{1}$, P. Caselli$^{2}$, A. Vasyunin$^{2,3}$, A. Palau$^{4}$}
\altaffiltext{1}{INAF-Osservatorio Astrofisico di Arcetri, L.go E. Fermi 5, Firenze, I-50125, Italy}
\altaffiltext{2}{Max-Planck-Institute for Extraterrestrial Physics, Giessenbachstrasse, D-85748 Garching, Germany}
\altaffiltext{3}{Ural Federal University, Ekaterinburg, Russia}
\altaffiltext{4}{Instituto de Radioastronom\'ia y Astrof\'isica, Universidad Nacional Aut\'onoma de M\'exico, PO Box 3-72, 58090 Morelia, Michoac\'an, M\'exico}
%$^{4}$ UJF-Grenoble 1/CNRS-INSU, Institut de Plan\'etologie et d'Astrophysique de Grenoble (IPAG) UMR 5274, Grenoble, F-38041, France  \\

\date{Received - ; accepted -}

%\pagerange{\pageref{firstpage}--\pageref{lastpage}} \pubyear{2015}
%
%\maketitle
%
%\label{firstpage}

\begin{abstract}
Phosphorus is a crucial element for the development of life, but
so far P-bearing molecules have been detected only in a few astrophysical
objects, hence its interstellar chemistry is almost totally unknown.
Here we show new detections of phosphorus nitride in a sample of
dense cores in different evolutionary stages of the intermediate- and
high-mass star formation process: starless, with protostellar 
objects, and with ultracompact \HII\ regions. All detected PN line widths 
are smaller than $\simeq 5$ \kms , and they arise from regions associated
with kinetic temperatures smaller than 100 K. Because the few previous
detections reported in the literature are associated with warmer and 
more turbulent sources, the results of this work show that PN can arise 
from relatively quiescent and cold gas. This information is challenging
for theoretical models that invoke either high desorption temperatures or
grain sputtering from shocks to release phosphorus into the gas phase.
Derived column densities are of the order of $10^{11-12}$ \cmq , 
marginally lower than the values derived in the few high-mass star forming
regions detected so far. 
New constraints on the abundance of phosphorus monoxide, the
fundamental unit of biologically relevant molecules, are also given.
\end{abstract}


\section{Introduction}
\label{intro}

Phosphorus is an important element for the development of life in the
Universe. It is one of the crucial components of nucleic acids, phospholipids, 
and the adenosine triphosphate (ATP) molecule, from which all forms of
life assume energy (e.g.~Pasek \& Lauretta~\citeyear{pel}). Despite this 
great relevance for biotic and pre-biotic chemistry, little is known about the 
phosphorus gas phase chemistry, and few measurements of its chemical 
compounds in the interstellar medium have been obtained so far.

Phosphorus is thought to be synthesized in massive stars and injected into 
the interstellar medium via supernova explosions (Koo et al.~\citeyear{koo}, 
Roederer et al.~\citeyear{roederer}). It has a low cosmic abundance relative 
to hydrogen ($\sim 2.8\times 10^{-7}$, Grevesse \& Sauval~\citeyear{ges}) lower 
than that of iron, magnesium, sodium, calcium, and aluminium, 
and it is thought to be depleted in the dense and cold interstellar medium
by a factor $\sim 600$ (e.g.~Wakelam \& Herbst~\citeyear{weh}).
Because P is essentially undepleted in the diffuse clouds 
(Lebouteiller et al.~\citeyear{leboutellier}), depletion of P should
be due to freeze-out onto the icy mantles of dust grains, and
its desorption mechanisms should be similar to those of all the other
icy mantle components.
Among the phosphorus-bearing molecules, the phosphorus nitride (PN)
is the first one detected in the interstellar medium towards three 
high-mass star forming regions: Orion (KL), Sgr B2, and W51,
in which the measured abundances are $\sim 1-4 \times 10^{-10}$, 
larger than theoretically expected from a pure low-temperature ion-molecule
chemical network (Turner \& Bally~\citeyear{teb}, Ziurys~\citeyear{ziurys}). 
Since then, it has been detected in few other high-mass dense cores (Turner et al.~\citeyear{turner}), 
as well as in the circumstellar material of carbon- and oxygen-rich 
stars (e.g.~Milam et al.~\citeyear{milam08}, De Beck et al.~\citeyear{debeck}) 
and tentatively in protostellar shocks (Yamaguchi~\citeyear{yamaguchi}). 
Other phosphorus-bearing molecules (e.g.~PO, CP, HCP, PH$_3$)
were detected in evolved stars (Tenenbaum et al.~\citeyear{tenenbaum}, 
De Beck et al.~\citeyear{debeck}, Ag\'undez et al.~\citeyear{agundez}), 
but never in dense star-forming cores so far. Due to this lack of observational
constraints, the chemistry of phosphorus in the interstellar medium is 
basically still unknown.

The few theoretical works focussed so far on the chemistry of interstellar 
phosphorus disagree in the prediction of the abundances of the main
P-bearing molecules. Charnley \& Millar (1994) indicate that PO,
the fundamental bond unit of many relevant biological molecules
(Maci\'a et al.~\citeyear{macia}), should have abundances similar 
to PN in hot molecular cores (up to $\sim 10^4$ yrs), 
while other molecules (e.g.~CP, HCP) would 
require formation timescales longer than the lifetime of hot cores, and hence 
should not be detectable in these environments. On the other hand, the theoretical 
predictions by Millar et al.~(\citeyear{millar}) and Adams et al.~(\citeyear{adams}) 
suggest that PN should be more abundant than PO by about two orders of
magnitude, while Thorne et al.~(\citeyear{thorne}) propose PO as the most 
abundant P-containing molecule based
on modelling and laboratory experiments. Therefore, the various
models predict different relative abundances of even the 
simplest molecules. Only observations, by testing and constraining 
the competing models, can advance the discussion.

In this work, we report on several new detections of PN in dense star-forming
cores where intermediate- and high-mass star formation is on-going. 
Moreover, new upper limits on the abundance of PO are given. 
We present the observations in Sect.~\ref{obs}.
Most of the targets belong to the sample of Fontani et al.~(\citeyear{fonta11}, 
hereafter F+11), who selected their objects 
based on these criteria: (i) cores relatively nearby (distance $\leq 5$ kpc)
not blended with others; (ii) cores in different evolutionary stages: 
starless cores without signs of embedded star formation activity (HMSCs), 
protostellar objects (HMPOs), and ultracompact \HII\ regions (UCHIIs). 
%We report also a new detection of PN in the well-known ultracompact \HII\ 
%region W51 ({\bf short description and references}). 
The results are described in Sect.~\ref{res} and discussed in Sect.~\ref{discu}.
%A summary of the main results of the work are given in Sect.~\ref{summary}.

\section{IRAM-30m telescope observations}
\label{obs}

{\it Run-1:} the PN (2--1) line, with a rest frequency of 93979.78 MHz, was 
observed towards the sources listed in Table~\ref{tab_sources} as part 
of the observations published in Fontani et al.~(\citeyear{fonta15a}) 
and Fontani et al.~(\citeyear{fonta15b}). A detailed description of this 
observing run is given in Sect. 2 of Fontani et al.~(\citeyear{fonta15a}).
The main observational parameters are summarised in Table~\ref{tab_lines}.

\vspace{0.8mm}
\noindent
{\it Run-2:} several transitions of the PO molecule were observed
in between 152656.98 and 152888.13~GHz from the 5th to the 
9th of June, 2015. The ones expected to be the brightest are reported 
in Table~\ref{tab_lines}. Calibration was performed following the chopper 
wheel technique (see Kutner \& Ulich~\citeyear{kutner}), with a calibration
uncertainty of up to $\sim 20 \%$. The spectra were obtained in
antenna temperature units, $T_{\rm A}^{*}$, and then converted to
main beam brightness temperature, $T_{\rm MB}$, via the relation
$T_{\rm A}^{*}=T_{\rm MB}\eta_{\rm MB}$ (where $\eta_{\rm MB}
=B_{\rm eff}/F_{\rm eff}$). 
The observations have been taken in wobbler-switching mode.
Pointing was checked every hour on nearby quasars or bright
\HII\ regions. Focus was checked at the beginning of each observing 
run and at sunrise, either on Saturn or on a nearby strong quasar.
The atmospheric conditions were very stable during the whole
observing shift, with a precipitable water vapour always in between 
$\sim 2$ and $\sim 7$ mm.

\vspace{0.8mm}
\noindent
For both runs, the main spectral and technical parameters are
listed in Table~\ref{tab_lines}. The spectral parameters (rest frequency, Einstein
coefficient, energy of the upper level) are taken from the Jet Propulsory Laboratory
catalog (Pickett et al.~1998).
All calibrated spectra were analyzed using the 
GILDAS\footnote{The GILDAS software is available at http://www.iram.fr/ IRAMFR/GILDAS} 
software developed at the IRAM and the Observatoire de Grenoble, and MADCUBAIJ
\footnote{Madrid Data Cube Analysis on ImageJ is a software
to visualize and analyze astronomical single spectra and datacubes 
(Mart\'in-Pintado et al., {\it in prep.}).}, developed
at the Center of Astrobiology (Madrid, INTA-CSIC).

% {\bf add observations towards W51?} VMR: NO

\section{Results}
\label{res}

\subsection{PN}
\label{pn}

\begin{figure}
\centerline{\includegraphics[angle=0,width=16cm]{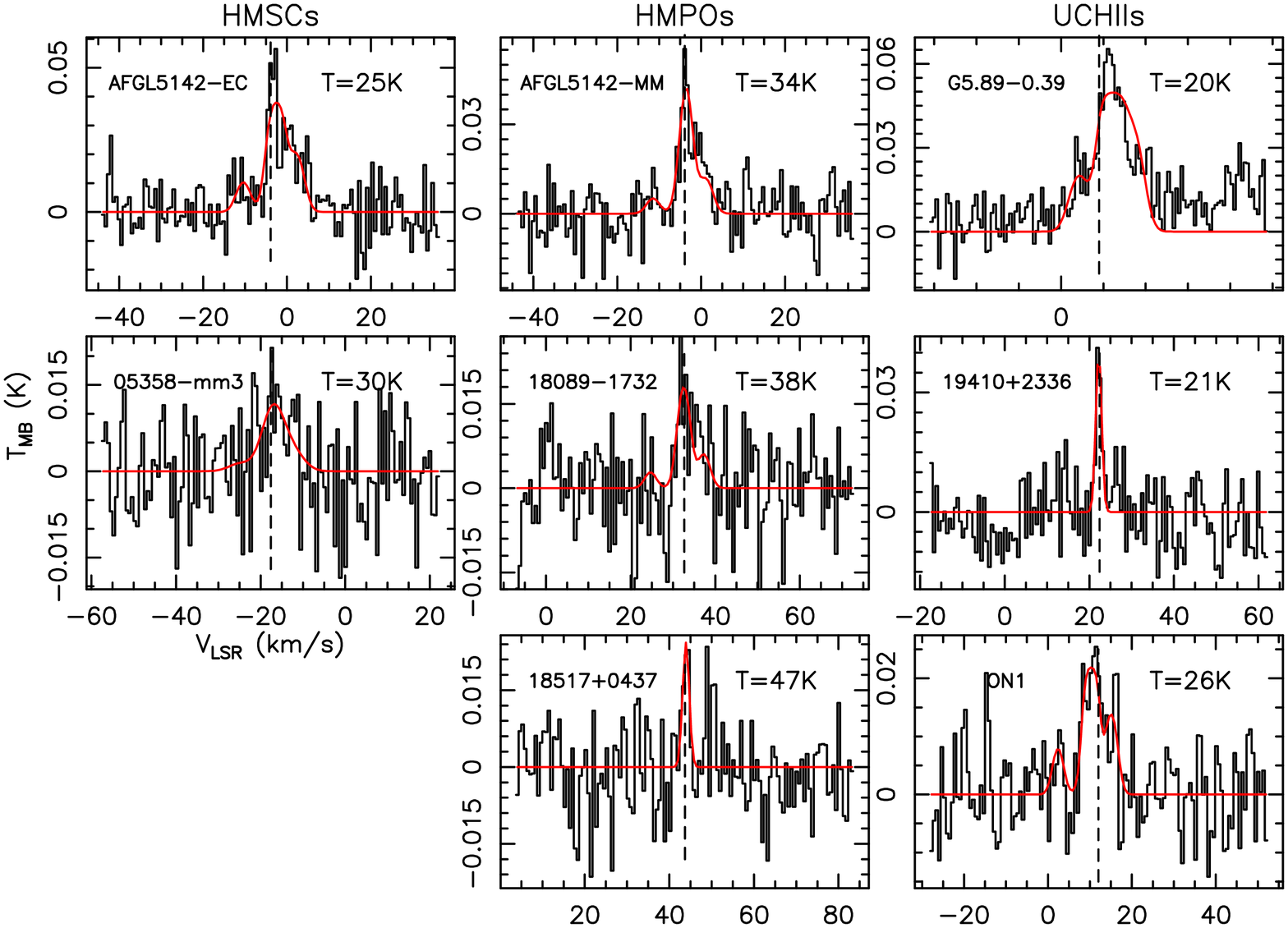}}
\caption[]{Spectra of the PN (2--1) lines detected towards the
cores of the F+11 survey. In each frame, the red line represents the
best fit to the hyperfine structure, except for 18517+0437 and 19410+2336,
for which Gaussian fits are shown. In each spectrum, the vertical dashed 
line indicates the core systemic velocity (Table~1 of F+11).
%Note that the two cores for which the $^{14}$N/$^{15}$N derived
%from the two molecules is not the same within the errors both
%belong to the same star forming region AFGL5142.
}
\label{PN_detections}
\end{figure}

In the sample of F+11 we have detected the PN (2--1) transition in two 
HMSCs, three HMPOs, and three UCHIIs. The two HMSCs are both 
defined as "warm" cores by F+11, based on their kinetic temperature 
larger than 20~K. These cores do not reveal signs of internal protostellar 
activity, but are located close to protostellar objects that can affect the gas 
temperature, and hence the gas chemistry (see F+11 for details). 
Moreover, the line detected towards the starless core AFGL5142--EC is
likely contaminated by the emission coming from the nearby protostellar
core AFGL5142--mm, given the large beam size ($\sim 27$\asec ).
The situation is different for the starless core 05358--mm3, because the
emission can be contaminated by the nearby HMPO 05358--mm1, but this
latter is undetected, thus we are confident that in the spectrum of 05358--mm3, 
the emission from the starless core is dominant. In Fig.~\ref{PN_detections} 
we show the PN (2--1) spectra of the eight detected sources.
The detections have been checked with the code MADCUBAIJ. The code 
provides synthetic spectra under Local Thermodynamic Equilibrium (LTE) 
conditions, taking into account the opacity of the lines. Thus, MADCUBAIJ was
used both to confirm the detections, and to complement the analysis performed 
with CLASS (see below). All the detected lines are consistent with the 
synthetic spectra produced assuming the temperatures and column densities 
of the sources. 
Moreover, we have carefully inspected the presence of possible
lines overlapping with PN (2--1) in the Cologne Database for Molecular
Spectroscopy\footnote{http://www.astro.uni-koeln.de/cdms/}, 
and ruled-out any contamination by nearby lines.
%{\bf Victor:please insert some lines of text to describe the program and the method, or
%give reference.}
%{\bf Victor: we should add here the results of W51 ... }

The lines of PN possess hyperfine structure due to the electric 
quadrupole moment of the $^{14}$N nucleus (see Cazzoli et 
al.~\citeyear{cazzoli}). The peak velocities and full width at half maximum
of the lines have been derived from fits that consider this structure.
The method has given good results for all sources except 18517+0437 
and 19410+2336, for which the lines are too narrow with respect to the
velocity resolution, so that the fit to the hyperfine structure provides line
parameters with too large uncertainties. For these, the lines have been fit with 
a single Gaussian. The results are given in Table~\ref{tab_sources}.
Derived line widths are usually larger than $\sim 3$ \kms . In F+11
all (but one) quiescent starless cores possess $\Delta {\rm v}$ of \H\ smaller
than 3 \kms , while all UCHIIs (but one) have $\Delta {\rm v} \geq 3$ \kms . Because
the UCHII regions are the sources in which shocks should have the most
significant influence on the kinematics, measured $\Delta {\rm v}$ of PN are
consistent with the idea that phosphorus needs 
a shock to be released from the dust grains and injected in the gas. 
However, in the two objects that have been fit with a single Gaussian 
(e.g.~18517+0437, 19410+2336) we have found 
relatively narrow lines ($\sim 1.6 - 1.8$ \kms). Moreover, for these lines
the simplified approach using a Gaussian gives an upper limit 
of the intrinsic value. These are the first PN lines with widths narrower 
than $\sim 5$ \kms , indicating that PN is not necessarily associated with 
shocks and can be found also in relatively quiescent gas. This point will 
be discussed better in Sect.~\ref{discu}.
Peak velocities are consistent within $\sim 1.5$ \kms\ 
with those of \H , \AMM , and NH$_2$D (F+11, Fontani et al.~\citeyear{fonta15a}),
except in G5.89--0.39, indicating that in general PN, \H , \AMM\ and 
NH$_2$D trace similar material. In G5.89--0.39, the clear velocity
difference indicates separate emitting regions to be investigated
at higher angular resolution.

The beam-averaged column densities have been derived from the 
integrated intensities (the sum of the intensities of all channels with signal) 
of the lines from Eq.~(A4) in Caselli et al.~(\citeyear{caselli}), assuming local 
thermodynamic equilibrium and optically thin conditions. 
All values are in between 1.6 and 10.4 $\times 10^{11}$ \cmq ,
smaller on average than the few previous observations in high-mass
star forming cores and dense clouds (i.e.~$\geq 10^{12}$ \cmq , Turner et al.~\citeyear{turner}).
Interestingly, the region NGC7538 was detected in PN (5--4) by
Turner et al.~(\citeyear{turner}) with the NRAO-12m telescope,
but it is undetected in the (2--1) line in this work. Note that the beam of 
NRAO-12m at $\sim 235$~GHz (rest frequency of PN(5--4)) is very similar 
to that of our observations. 
Therefore, beam dilution effects cannot explain the non-detection with the
IRAM-30m Telescope. Assuming a core size of 22\asec\ and a gas temperature 
of 100 K, they derived a PN column density of 5.8$\times 10^{12}$ \cmq . 
This value is larger than the upper limit derived in this work (2.2$\times 10^{11}$ \cmq ),
in which, however, the assumed \Tex\ is 20 K. Assuming \Tex = 100 K and a source
size of 22\asec , we obtain $\sim 2\times 10^{12}$ \cmq , more consistent
with the estimate of Turner et al.~(\citeyear{turner}) within the uncertainties.
The integrated areas, the excitation temperatures (\Tex ), and the
corresponding column densities of PN are listed in Table~\ref{tab_sources}.
The analysis using MADCUBAIJ provides very similar abundances
(see Col.~7 of Table~\ref{tab_sources}), and gives low opacities 
($<$0.005), supporting the assumption that the lines are optically thin.
This opacity is calculated assuming that the source fills the beam,
which is not a good approximation for our sources. 
However, even assuming a source size of 5\asec , a realistic average
angular size for our sources (see F+11), the opacities are still below $\sim$0.05, 
perfectly consistent with the assumption of optically thin conditions.
Unfortunately, the PN abundance cannot be derived for all objects because 
data of the thermal dust (sub-)mm continuum emission, from which the
H$_2$ column density can be obtained, are still lacking.

Because for optically thin lines \Tex\ cannot be derived directly from 
the spectrum, we have assumed the gas kinetic temperatures reported
in Table~A.3 of F+11. All spectral parameters utilised in the derivation
of the column densities have been taken from the Jet Propulsion Laboratory 
(Pickett et al.~1998) catalogue. The assumption of optically thin conditions
is consistent with the moderate opacity ($\tau \sim 1$ or lower) derived in 
the sources in which $\tau$ can be computed directly from the measure of
the relative intensities of the hyperfine components (for details, see the CLASS 
manual\footnote{https://www.iram.fr/IRAMFR/GILDAS/doc/html/class-html/class.html}).
In two cases, AFGL5142--EC and ON1, the fit to the hyperfine structure
provides $\tau \geq 3$, but the uncertainty on $\tau$ is comparable to the value,
and this, together with the low optical depth derived with MADCUBAIJ,
prevents us to assume optically thick conditions. Therefore, we have decided
to conservatively assume optically thin conditions in these two cases as well.
For undetected sources, the upper limit on the integrated
intensity has been calculated from the $3 \sigma$ rms of the spectrum
from the formula $\int T_{\rm MB}dv={\rm 3 \sigma}\Delta{\rm v}/(2\sqrt{\ln{2}/\pi})$, 
valid for a Gaussian profile with peak temperature equal to the $3 \sigma$ rms.
%{\bf Victor, about the consistency of the upper limits: it is true that perhaps
%the two methods give different (slightly) values, but the upper limits on PN
%should also be consistent with the N(PN) in the detected source, so I would
%use the same method for detections and non-detections. And I would stick
%on the method used currently because it is the same I used for CN, N2H+ , and
%all other molecules with which I make column density ratios in the following.}

%We are confident that this simplified approach gives a reasonable upper limit
%because in the detected sources the integrated intensity and the area derived
%from a Gaussian fit are different of $\sim 8 \%$ at most, i.e. well within the 
%calibration uncertainties.

\begin{table*}
\begin{center}
\caption[] {Observed and derived parameters: $v_{\rm peak}$ and $\Delta {\rm v}$
are peak velocity in the Local Standard of Rest and full width at half maximum 
of the PN (2--1) lines derived from the fitting procedure used into CLASS described 
in Sect.~\ref{pn}, except when specified differently (see footnotes); 
$\int T_{\rm MB}dv$ is the total integrated intensity of the line; \Tex\ is the 
excitation temperature, assumed equal to the kinetic temperature given by F+11;
$N$(PN), $N$(PO), and $N$(CN) are the total beam-averaged column 
densities of PN, PO (this work) and CN (Fontani et al.~\citeyear{fonta15b}).
Uncertainties are derived from the propagation of errors, and do not include
the calibration error (of the order of $10 \%$).}
\label{tab_sources}
\scriptsize
\begin{tabular}{ccccccccc}
\hline \hline
source & $v_{\rm peak}$ & $\Delta {\rm v}$ & $\int T_{\rm MB}dv$ & \Tex\ & $N$(PN) & $N$(PN)$^{b}$ & $N$(PO)$^{b}$ & $N$(CN)$^{c}$ \\
             & (\kms )   &  (\kms )    & K \kms\     & K & $\times 10^{11}$ \cmq\ & $\times 10^{11}$ \cmq\ & $\times 10^{11}$ \cmq\ & $\times 10^{14}$ \cmq\ \\
\hline
\cline{1-9}
\multicolumn{9}{c}{HMSC}   \\
00117--MM2    &   -- & 1.6$^{a}$ &  0.032 & 14  & $< 0.6$ & &  & \\
AFGL5142--EC &   --2.63  &  3.8(0.6)  &   0.32(0.02) &  25  & 8.3(0.4) & 5.8 & $< 20$ & 4.9(0.3) \\
05358--mm3   &  --17.52  &  6(2)  &  0.11(0.01)  &  30  &  2.9(0.4) & 10 & $< 13$  & 2.4(0.1) \\
G034--G2     &  --  & 1.6$^{a}$ &  0.037 &   16  & $< 0.8$ & & \\
G034--F2      & --   & 1.6$^{a}$ &  0.037 &   16  & $< 0.8$ & & \\
G034--F1      & --   & 1.6$^{a}$ &  0.043 &   16  & $< 0.9$  & & \\
G028--C1     & --   & 1.6$^{a}$ &   0.041 &  17  & $< 0.8$  & & \\
I20293--WC  & --  & 1.6$^{a}$ &   0.036 &  17  & $< 0.7$  & & \\
I22134--G      & --  & 1.6$^{a}$  &  0.036 &  25  & $< 0.9$  & & \\
I22134--B      & --  & 1.6$^{a}$  &  0.036 &	17   & $< 0.7$ & & \\
\cline{1-9}
\multicolumn{9}{c}{HMPO}   \\
00117--MM1   & -- & 2.9$^{a}$  &  0.07 &  20  & $< 1.5$ & & \\
AFGL5142--MM  & --3.66 & 3.7(0.6) &  0.21(0.02) &  34 &  6.9(0.6) & 12 & $< 12$ & 4.3(0.2) \\
05358--mm1      & --  & 2.9$^{a}$  &  0.09  & 	39  & $< 3.3$ & &  \\
18089--1732   & 33.04  & 3(1) &   0.09(0.02) &     38  &  3.3(0.6) & 6.3 & $< 6.3$ & 6.3(0.2) \\
18517+0437$^{d}$    & 44.12  &  $<$1.8(0.7)$^{d}$ &  0.05(0.02)$^{d}$ &    47  & 2.0(0.7) & 4.4 & $< 4.4$ & 3.1(0.2) \\
G75--core        &  --  & 2.9$^{a}$ & 0.07     &     96  & $< 5.8$ & & \\
I20293--MM1    & --  & 2.9$^{a}$ &  0.09   &     37  & $< 3.2$ & & \\
I21307              & --  & 2.9$^{a}$  &  0.07    &     21   & $< 1.6$ & & \\
I23385             & --   & 2.9$^{a}$  &  0.05    &     37   & $< 1.7$ & & \\
\cline{1-9}
\multicolumn{9}{c}{UCHII}   \\					       
G5.89--0.39  &  12.0  &  5.2(0.8) & 0.5(0.3) &   20  & 10.4(0.8) & 14.5 & $< 14$ & 15(4) \\
19035--VLA1 &  --     &  3.5$^{a}$  &  0.15   &   39   & $< 5.6$ & & \\
19410+2336$^{d}$  &  22.5 & $< $1.6(0.4)$^{d}$  & 0.07(0.01)$^{d}$  &  21 & 1.6(0.3) & 3.3 & $< 3.3$ & 10(3) \\
ON1                &  12.0 & 2.9(0.5) &  0.14(0.01)	  &  26  & 3.8(0.04) & 9.5 & $< 6.3$ & 5(2) \\
I22134--VLA1  &  --   & 3.5$^{a}$  &  0.075    &  47  & $< 3.1$ & & \\ 
23033+5951   &  --    & 3.5$^{a}$  & 0.15     &  25  & $< 3.9$ & & \\
NGC7538--IRS9  &  -- & 3.5$^{a}$ &  0.10  &   20  & $< 2.2$ & & \\
\hline
\end{tabular}
\end{center}
\begin{flushleft}
$^{a}$ fixed line width assumed to derive the upper limit on $\int T_{\rm MB}dv$
(see Sect.~\ref{pn}). The value is the average line width derived from
the detected sources for the HMPOs and the UCHIIs, while for the HMSCs we
have taken the average line width of the \D\ (2--1) line (F+11); \\
$^{b}$ derived from MADCUBAIJ (see Sects.~\ref{pn} and ~\ref{po}).
Upper limits on PO are given only for the sources detected in PN;\\
$^{c}$ from Fontani et al.~(\citeyear{fonta15b}); \\
$^{d}$ values derived from Gaussian fits to the lines.\\
\end{flushleft}
\end{table*}
\normalsize

\subsection{PO}
\label{po}

PO lines are not detected in any source, but we have computed
upper limits on the column densities with the code MADCUBAIJ.
The upper limits for the integrated intensity have been derived 
from the formula $3\sigma\times\Delta {\rm v}/\sqrt{N_{\rm chan}}$, where 
$N_{\rm chan}$ is the number of channels covered by the linewidth $\Delta$v.
The upper limits of PO for the objects detected in PN are reported in 
Table~\ref{tab_sources}. They are slightly higher or of the order of the PN 
column densities. Although for PO we only have upper limits, our
findings are not against the results found in the evolved 
oxygen-rich stars VY CMa (Tenenbaum et al.~\citeyear{tenenbaum}) 
and IK Tau (De Beck et al.~\citeyear{debeck}), detected in both PN and PO, 
in which their abundances are of the same order of magnitude as well. 
%{\bf Victor: please add details here.}

\begin{table*}
\begin{center}
\caption[] {Observed transitions and technical parameters}
\label{tab_lines}
\scriptsize
\begin{tabular}{llccccccc}
\hline \hline
molecular line & line rest frequency & $Log_{10}(A_{\rm ij})$ & $E_{\rm u}$ & HPBW  & $\Delta v$ & $T_{\rm sys}$ & $\eta_{\rm MB}$ \\
  & (GHz) & & (K) & (\asec ) & (\kms ) & K &  \\
\cline{1-8}
PN J=2--1 & 93.97978 & $-4.53516$ & 6.8 & $27$ & 0.62 & $\sim 100 - 120$ &  0.84 \\
\cline{1-8}
PO J=7/2-5/2 $\Omega =1/2$,F=4-3,l=e & 152.65698 & $-4.75076$    & 15.7 & 16 & 0.096 & $\sim 200 - 350$ & 0.65/0.93 \\
PO J=7/2-5/2 $\Omega =1/2$,F=3-2,l=e & 152.68028 & $-4.77193$    & 15.7 & 16 & 0.096 & $\sim 200 - 350$ & 0.65/0.93 \\
PO J=7/2-5/2 $\Omega =1/2$,F=4-3,l=f  & 152.85545 & $-4.74923$    & 15.7 & 16 & 0.096 & $\sim 200 - 350$ & 0.65/0.93 \\
PO J=7/2-5/2 $\Omega =1/2$,F=3-2,l=f  & 152.88813 & $-4.77033$    & 15.7 & 16 & 0.096 & $\sim 200 - 350$ & 0.65/0.93 \\
\hline
\end{tabular}
%\tablefoottext{b}{Telescope HPBW at the central frequency of the spectral window.}
%\tablefoottext{c}{Maximum spectral resolution obtained with FTS.}
\end{center}
\begin{flushleft}
$^{a}$ Total spectral window covered by the FTS correlator.
\end{flushleft}
\end{table*}
\normalsize

\section{Discussion and conclusions}
\label{discu}

\begin{figure}
\centerline{\includegraphics[angle=0,width=14cm]{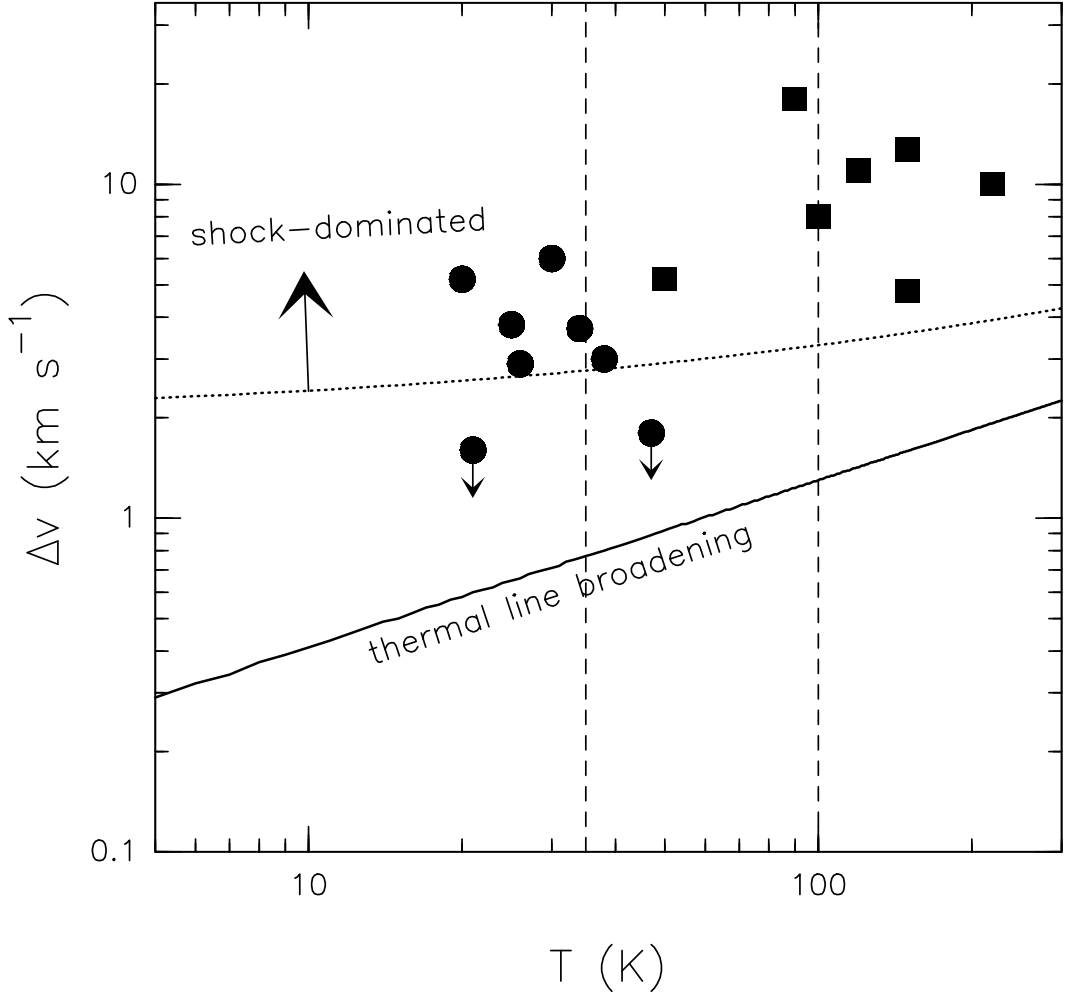}}
\caption[]{Line widths derived from the PN (2--1) line in this work (circles)
and in Turner et al. ~(\citeyear{turner}, squares) against the gas kinetic
temperature. The vertical dashed lines indicate the desorption temperatures 
predicted by the theoretical model of Charnley \& Millar~(\citeyear{charnley}), 
i.e. 100$\pm15$ K, and the possible smaller one derived from 
Garrod \& Herbst~(\citeyear{geh}) of $\sim 35$ K. The expected thermal 
line broadening, $\Delta v_{\rm th}$, is indicated by a solid oblique line.
Finally, the dotted line marks an arbitrary threshold of $\Delta v_{\rm non-th}\simeq 2$ \kms\
(where $\Delta v_{\rm non-th}$ represents the non thermal component of the
line), corresponding to about ten times $\Delta v_ {\rm th}$, 
in which the internal motions likely start to be affected by shock-induced 
turbulence. The two circles with arrows indicate 18517+0437 and 19410+2336,
for which the line width was derived from a Gaussian fit, hence it represents
an upper limit to the intrinsic line width.
%Note that the two cores for which the $^{14}$N/$^{15}$N derived
%from the two molecules is not the same within the errors both
%belong to the same star forming region AFGL5142.
}
\label{fig_victor}
\end{figure}

Before this work, PN was detected in six hot ($T\geq 50$ K) and turbulent 
($\Delta v \geq 6$ \kms) high-mass star forming cores
(Turner \& Bally~\citeyear{teb}, Ziurys~\citeyear{ziurys}, Turner et al.~\citeyear{turner}).
In this work we have detected eight regions with line widths narrower than
6 \kms , and temperatures in the range $\sim 20 - 60$ K, which suggests that
PN can also be formed in relatively cold and quiescent material. The comparison 
between our new detections and the previous ones are shown in Fig.~\ref{fig_victor}.
These results are challenging for chemical models that explain the formation of PN via 
thermal desorption of PH$_3$ from grain mantles at temperature above 
$\sim 100$ K (Charnley \& Millar~\citeyear{charnley}, solid vertical line in 
Fig.~\ref{fig_victor}), followed by rapid gas phase reactions ($10^4$ yrs) which 
transform it into PN, PO, or atomic P. They also disagree with previous observations 
that claimed high depletion of P (depletion factors of $\sim 10^3$) in dense star 
forming cores, suggesting violent mechanisms like grain disruption to have a 
significant amount of phosphorus in the gas phase (Turner \& Bally~\citeyear{teb}). 
In fact, as shown in Sect.~\ref{pn}, some lines are narrower than $\sim 2$ \kms ,
hence the non-thermal motions are very likely not dominated by shocks 
(Fig.~\ref{fig_victor}).
A lower sublimation temperature of PN ($\sim 35 - 40$ K) 
has been claimed (see e.g.~Garrod \& Herbst~\citeyear{geh}), which however 
is still higher than the kinetic temperature measured in many sources detected
in this work (see Fig.~\ref{fig_victor}).

We have not found any (anti-)correlation between line parameters and
other core properties (kinetic temperature, line widths of other molecules, deuterium fractionation), 
except for a correlation between line width of PN and kinetic temperatures 
(Fig.~\ref{fig_victor}), which was expected since the warmer cores are
typically more turbulent. Interestingly, the PN line widths in HMSCs are 
larger than those measured in CN (Fontani et al.~\citeyear{fonta15b}), 
suggesting different emitting regions,
%In particular, PN could
%trace either a more extended envelope, possibly shocked by external triggers,
%or internal material influenced by incipient star formation, 
but any interpretation cannot be supported without high angular resolution maps
of both PN and CN.

\begin{figure}
\centerline{\includegraphics[angle=90,width=15cm]{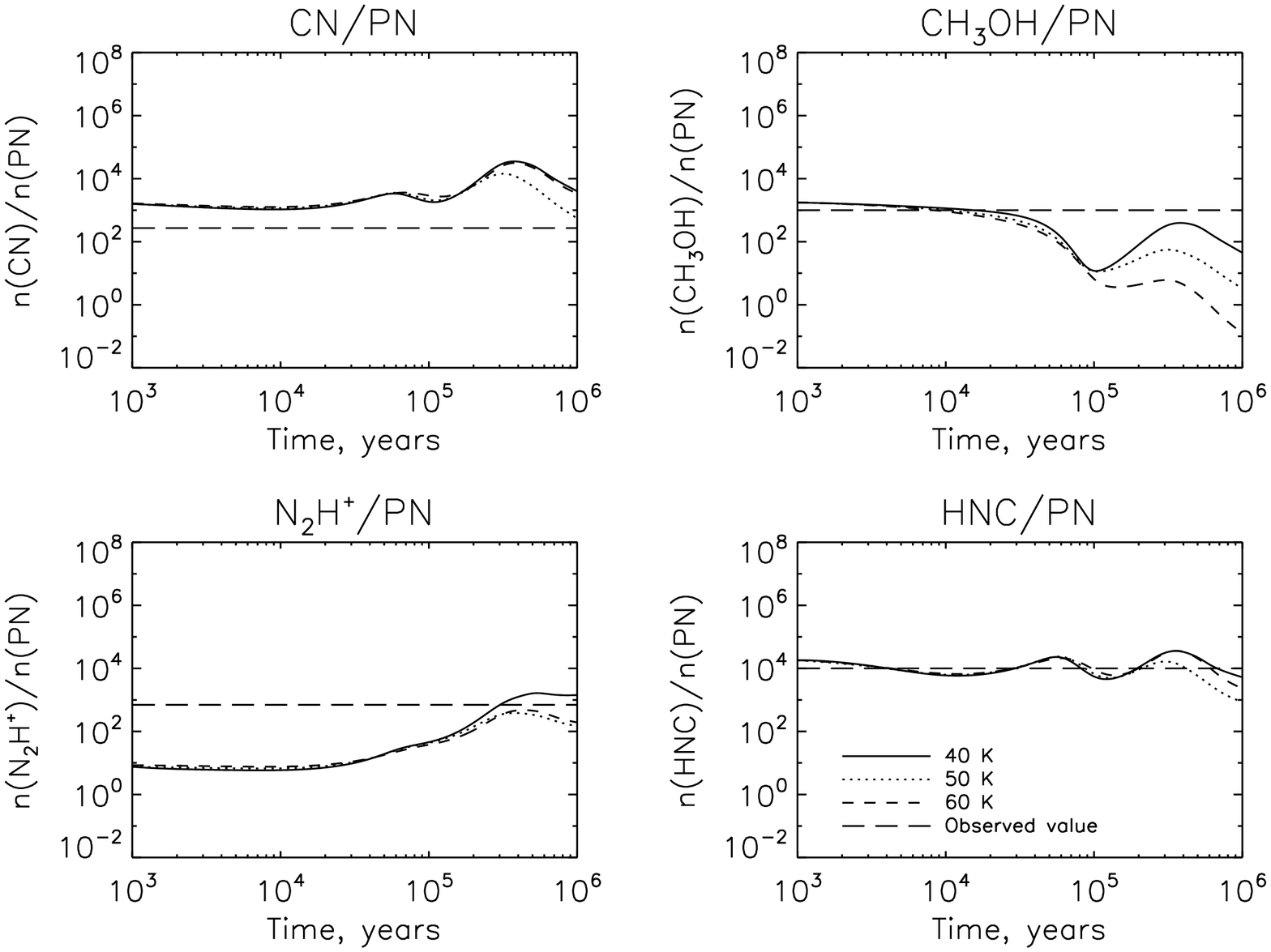}}
\caption{Volume density ratios as a function of time of the species indicated on top 
of each panel, calculated from the chemical models described in Sect.~\ref{discu}.
Straight, dotted, and dashed curves represent a constant temperature of 40,
50, and 60~K, respectively. The dashed line in each frame indicates the observed
average density ratios (see Sect.~\ref{discu}).}
\label{models}
\end{figure}

To model the abundance ratios, we utilized chemical model based on 
Vasyunin \& Herbst~\citeyear{veh}). We adopted "low metals" initial chemical 
composition (EA1, Table~1 in Wakelam \& Herbst~\citeyear{weh}) to account for the depletion 
of elemental phosphorus in cold interstellar medium.  The chemical evolution of a cloud 
was modeled in two stages. First, chemical evolution was calculated for a cold dark 
clump with density $n$(H$_2$)=$10^4$ \cmc , visual extinction $A_v=100$ mag, and 
temperature $T=10$ K for $10^6$ years. After that, gas density and temperature were 
increased to $n$(H$_2$)=$10^5$ \cmc , and $T=40 (50, 60)$ K, correspondingly, and 
chemical evolution has been followed for another $10^6$ years. The resulting abundance 
ratios for this time span are showed on Fig.~\ref{models}, which shows the predictions 
for CN/PN, \METH /PN, \H /PN, and HNC/PN.

The column densities of the species aforementioned have been measured 
in all the sources detected in PN (Fontani et al.~\citeyear{fonta15a}, Fontani 
et al.~\citeyear{fonta15b}, Fontani et al.~\citeyear{fonta14}). Therefore, we have 
compared the volume density ratios observed to the values predicted by the models, 
assuming that the emitting diameter is the same for the different molecules. 
We find that the average CN/PN ratio is $\sim 270$. 
This has been obtained correcting the beam-averaged N(CN) listed in Table 2 
of Fontani et al.~(\citeyear{fonta15b}) for the factor $(11/26)^2$ to take the
different beams into account. With a similar approach, we find an average \H /PN$\sim 700$,
an average \METH /PN$\sim 1000$, and an average HNC/PN of $\sim 10^4$. 
These latters have been derived from the column densities of \H\ and \METH\ given 
in Fontani et al.~(\citeyear{fonta15b}), Fontani et al.~(\citeyear{fonta15a}), and
Fontani et al.~(\citeyear{fonta14}), respectively. Inspection of Fig.~\ref{models} 
tells us that the ratios \H /PN, \METH /PN, and HNC/PN are globally in agreement
with the models that assume an age of the objects $\sim 10^5 - 10^6$ yrs
and $T=40$ K, mainly constrained by \METH /PN. On the other hand, 
the CN/PN ratio cannot be reproduced by any model. 
A possible explanation can be a significantly different emitting region 
of the two molecules. In fact, in the observations of Fontani et al.~(\citeyear{fonta15b}),
the beam is 11\asec , but if the CN emission arises from a more extended area, 
then the column density given by Fontani et al.~(\citeyear{fonta15b}) is a lower
limit. That CN is likely a tracer of the more extended envelope rather than the
densest portions of the sources is consistent with the finding that its abundance
increases with the presence of UV photons (e.g.~Fuente et al.~\citeyear{fuente},
Palau et al.~\citeyear{palau07}).
%This could explain why CN/PN in the models is larger than in the observations.
However, to test this scenario, and to better understand the origin of PN in dense
star-forming cores, future higher angular resolution observations are absolutely
required. 

\section*{Acknowledgments}

The authors are grateful to the IRAM staff for its help during
the observations of the IRAM-30m data. This work was partly supported 
by the Italian Ministero dellÕIstruzione, Universit\'a e Ricerca through the 
grant Progetti Premiali 2012 - iALMA. A.P. acknowledges 
financial support from UNAM-DGAPA-PAPIIT IA102815 grant, M\'exico.
P.C. and A.V. acknowledge support from the European Research
Council.
%Dr ???for a critical reading of the original version of the
%paper and an anonymous referee for very useful comments that improved
%the presentation of the paper.

{}

\end{document}